\documentclass{kapproc} 

\usepackage{procps} 

\usepackage[dvips]{graphicx}

\upperandlowercase

 \let\footnote\savefootnote

\normallatexbib 

\begin{document}

\articletitle
{Absorption Line Studies in the Halo}

\chaptitlerunninghead{Absorption line studies} 

\author{Philipp Richter}

\affil{
Osservatorio Astrofisico di Arcetri, Florence, Italy,\\
Institut f\"ur Astrophysik, IAEF, Bonn, Germany}

\begin{abstract}

Significant progress has been made over the last few years to
explore the gaseous halo of the Milky Way by way of absorption
spectroscopy. I review recent results on absorption line studies
in the halo using various instruments, such as the 
{\it Far Ultraviolet Spectroscopic Explorer}, the
{\it Space Telescope Imaging Spectrograph}, and
others. The new studies imply that the infall of low-metallicity
gas, the interaction with the Magellanic Clouds, and the
Galactic Fountain are responsible for the phenomenon of the
intermediate- and high-velocity clouds in the halo. 
New measurements of highly-ionized gas in the
vicinity of the Milky Way indicate that these 
clouds are embedded in a corona of hot gas that extends
deep into the intergalactic space.

\end{abstract}

\section{Introduction}

Studying the gaseous halos of galaxies is important
to understand the various complex processes that balance
the exchange of gaseous matter and energy between
individual galaxies and the intergalactic medium.
Supernova explosions in spiral galaxies
create large cavities filled with hot gas in the gaseous disk -
such gas eventually breaks out of the the disk and flows into
the halo where part of it can cool and condense before falling
back onto the disk. This process - also called
``Galactic Fountain'' (Shapiro \& Field 1976) - was for
many years thought to be responsible for the
phenomenon of the ``High-Velocity Clouds'' (HVCs) in
our Galaxy. Also interaction and merging of galaxies 
will transport large amounts of interstellar
material into the halos and immediate intergalactic environment
of galaxies. Finally, left-over gas from the formation
of galaxies and galaxy groups is expected to contribute
to the circumgalactic medium around galaxies in the
low-redshift Universe.
To understand the intergalactic gaseous environment and the halos of
galaxies it is therefore important to 
study each of these various components in detail.

With the availability of space-based spectroscopic
instruments operating in the UV and FUV, such as the
the {\it Orbiting and Retrievable Far and Extreme
Ultraviolet Spectrometers} (ORFEUS), the {\it
Space Telescope Imaging Spectrograph} (STIS), and
the {\it Far Ultraviolet Spectroscopic
Explorer} (FUSE) it has become possible to
explore the gaseous Milky Way halo and the very local
intergalactic medium 
in absorption against distant extragalactic
UV background sources like quasars (QSOs) and
Active Galactic Nuclei (AGNs). The UV and FUV spectroscopic
range is particularly interesting for studying the
low-density, multiphase circumgalactic medium, because
many atomic and molecular species and their
ions have their electronic transitions in the
region between 900 and 1800 \AA\,
(e.g., H$_2$, H\,{\sc i}, D\,{\sc i}, N\,{\sc i}, 
O\,{\sc i}, Si\,{\sc ii}, Fe\,{\sc ii}, C\,{\sc iv} and
O\,{\sc vi}). 
Measurements of absorption lines from these
species therefore allow us to analyze in detail the gas
in the halo of the Milky Way in all of its
phases (i.e., from molecular to highly-ionized).

\section{Intermediate- and High-Velocity Clouds}

The origin of the Intermediate- and High-Velocity Clouds
(IVCs and HVCs, respectively) in the Milky Way halo 
has been controversial for
a long time. These neutral gas clouds are observed in
H\,{\sc i} 21cm emission at radial velocities that
deviate substantially from Galactic rotation models
(e.g., Wakker \& van\,Woerden 1997). 
However, unveiling the nature of the IVCs and
HVCs is a difficult task.
For most of the IVCs and HVCs it is impossible
to directly derive distances. Therefore, the most
valuable information about their origin comes 
from metal-abundance studies using FUV absorption
spectroscopy. An extensive summary of many
absorption line measurements of Galactic halo clouds
is provided by Wakker (2001). 
Using high signal-to-noise UV and FUV
absorption line data from STIS and FUSE it
has become possible in the last few years 
do reliably determine metal abundances and physical
conditions in several intermediate- and high-velocity clouds.

For one of the most prominent HVCs,
Complex C, several studies using FUSE and STIS
data imply metallicities varying between 
$0.1$ and $0.3$ solar along different lines
of sight (Wakker et al.\,1999; Richter et al.\,2001a;
Collins et al.\,2003; Tripp et al.\,2003). Complex C
also exhibits a 
notable underabundance of nitrogen.
Probably, this HVC therefore repesents a metal-deficient
intergalactic gas cloud that is falling onto
the Milky Way, and that recently has started to mix
with outflowing (metal-rich) Galactic Fountain gas. An
example for absorption of neutral oxygen in
Complex C is presented in Fig.\,1.
Another prominent HVC complex for which accurate abundances
have been measured is the Magellanic Stream. It 
has abundances close to those of the Small Magellanic Cloud
(SMC) and thus likely represents material stripped
out of the SMC during a close encounter with
the Milky Way (Lu et al.\,1998; Sembach et al.\,2001).
In contrast to the HVCs, absorption line 
studies of several IVCs show that
these clouds tend to have higher abudances, close
to those found in the local ISM (Richter et al.\,2001a;
Richter et al.\,2001b). It thus appears
plausible that the IVCs 
represent the return flow 
of a Galactic Fountain (Shapiro \& Field 1976).
These clouds also exhibit small and dense 
filamentary substructure containing molecular
hydrogen, H$_2$, and dust.
(Richter et al.\,2003; Richter, Sembach \&
Howk 2003). 

The recent absorption line measurements of IVCs and HVCs
clearly show that various different processes are responsible
for the phenomenon of high-velocity neutral gas clouds in the
halo of the Milky Way - they cannot have a single origin.

\begin{figure}[t!]
\centerline{\includegraphics[width=0.8\textwidth]{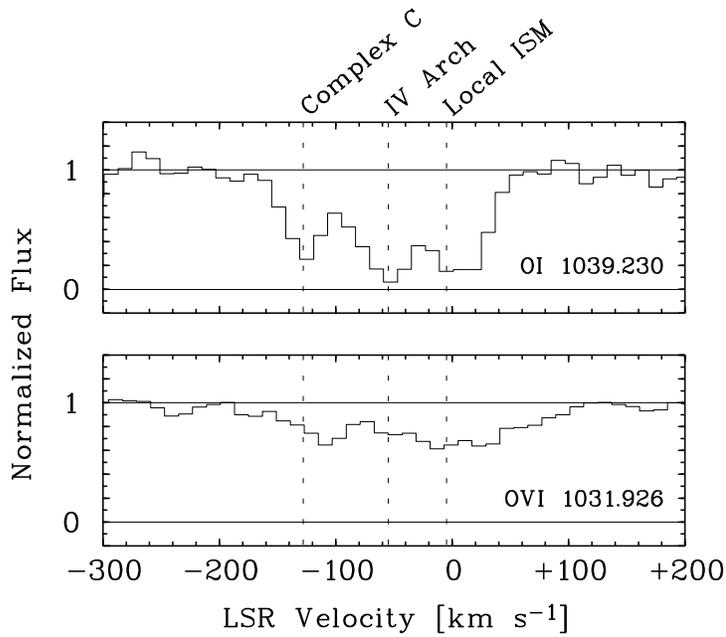}}
\caption{Interstellar absorption of neutral oxygen (O\,{\sc i}) 
and highly-ionized oxygen (O\,{\sc vi}) in the direction of the 
quasar PG\,1259+593 ($l=120.6$, $b=+58.1$, $z_{\rm em}=0.478$).
The O\,{\sc i} absorption shows two Galactic halo components 
at $-128$ km\,s$^{-1}$ (high-velocity cloud Complex C) and
at $-55$ km\,s$^{-1}$ (intermediate-velocity cloud IV Arch), while
local ISM absorption is seen at $-5$ km\,s$^{-1}$. These three
main velocity components are marked with dotted lines. Also
the O\,{\sc vi} absorption extends to high negative velocities,
possibly arsising in the interface regions between the neutral
HVC gas (as sampled by O\,{\sc i}) and a surrounding hot
medium - the Galactic Corona (see also Fox et al.\,2003).}
\end{figure}

\section{The Galactic Corona}

It was Lyman Spitzer (1956) who proposed the existence
of a hot gaseous corona around the Milky Way that
would confine the cooler and denser IVCs and HVCs with its
thermal pressure (see also Spitzer 1990; McKee 1993).
Absorption line spectroscopy in the
UV and FUV is well suited to study hot gas
in the vicinity of the Milky Way, as this range 
contains a number of lines from highly-ionized species,
such as C\,{\sc iv}, N\,{\sc v}, and O\,{\sc vi}. These lines
sample gas in the temperature range between 
$1\times 10^5$ and $5\times 10^5$ K. Studying O\,{\sc vi}
absorption in the halo is particularly interesting,
as O\,{\sc vi} has the highest ionization
potential of the three ions listed above.

The first absorption-line studies of the Galactic Corona were
based on observations of Si\,{\sc iv}, C\,{\sc iv}, and
N\,{\sc v} obtained with the {\it International 
Ultraviolet Explorer} (IUE; Savage \& de\,Boer 1979) 
and the {\it Hubble Space
Telescope} (HST; Savage, Sembach, \& Lu 1997).
Using data from the {\it Orbiting
and Retrievable Far and Extreme Ultraviolet Spectrometers}
(ORFEUS), Widmann et al.\,(1998) presented the 
first systematic study of O\,{\sc vi} absorption 
in the halo. With the availability of a large 
number of FUSE absorption spectra from extragalactic
background sources our knowledge about the 
$\sim 10^5$ K gas component in the halo
(as traced by O\,{\sc vi} absorption) has further improved
during the last few years.
Wakker et al.\,(2003), Savage et al.\,(2003), and
Sembach et al.\,(2003) present a large survey of
O\,{\sc vi} absorption along 102 lines of sight 
through the Milky Way halo. They find strong 
O\,{\sc vi} absorption in a radial-velocity range from 
approximately $-100$ to $+100$ km\,s$^{-1}$
with logarithmic O\,{\sc vi} column densities ranging from 
$13.85$ to $14.78$ (Savage et al.\,2003). 
At these radial velocities, the  O\,{\sc vi} absorbing
gas should be located in the thick disk and/or halo
of the Milky Way.
The distribution of the 
O\,{\sc vi} absorbing gas in the thick disk and halo 
is not uniform, put appears
to be quite irregular and patchy. 
A simple model assuming 
a symmetrical plane-parallel
patchy layer of O\,{\sc vi} absorbing material 
provides a rough estimate for the exponential O\,{\sc vi}
scale height in the halo. Savage et al.\,(2003) 
find $h_{\rm O\,VI}\sim2.3$ kpc with an $\sim 0.25$ dex
excess of O\,{\sc vi} in the northern Galactic polar region.
The correlation of O\,{\sc vi} with other ISM tracers 
such as soft X-ray emission, H$\alpha$, and H\,{\sc i}
21cm is rather poor (Savage et al.\,2003).  
Mixing of warm and hot gas and radiative 
cooling of outflowing hot gas from
supernova explosions in the disk could explain
the irregular distribution of O\,{\sc vi}
absorbing gas in the halo of the Milky Way.

\section{Local Group Gas}

O\,{\sc vi} absorption towards extragalactic background
sources is seen not only at radial velocities 
$|v_r|\leq100$ km\,s$^{-1}$, but also at 
higher velocities (Wakker et al.\,2003; Sembach et al.\,2003).
Studying high-velocity O\,{\sc vi} is of crucial 
interest to understand the immediate intergalactic 
environment of the Milky Way
and the various
processes that determine the distribution of hot
gas in in the Local Group. 

From their survey of high-velocity O\,{\sc vi} absorption
Sembach et al.\,(2003) find that probably more than 60 percent
of the sky at high velocities is covered by ionized hydrogen 
(associated with the O\,{\sc vi} absorbing gas) above
a column density level of log $N$(H$^+)=18$, assuming
a metallicity of the gas of $0.2$ solar.
Some of the high-velocity O\,{\sc vi} detected with FUSE 
appears to be
associated with known high-velocity H\,{\sc i} 21cm
structures (e.g., the High-Velocity Clouds Complex A,
Complex C, the Magellanic Stream, and the Outer Arm).
Other high-velocity O\,{\sc vi} features, however, have no counterparts
in neutral gas. The high radial velocities for most of these
O\,{\sc vi} absorbers are incompatible with those expected
for Galactic halo gas (even if the halo gas motion is
decoupled from the underlying rotating disk). A transformation
from the LSR into the GSR and LGSR velocity reference frames
reduces the dispersion about the mean of the high-velocity
O\,{\sc vi} centroids (Sembach et al.\,2003). This can be
interpreted as evidence, that {\it some} of the O\,{\sc vi} high-velocity
absorbers are intergalactic clouds in the Local Group
rather than clouds directly associated with the Milky Way.
The presence of local intergalactic O\,{\sc vi} absorbing gas
is in line with
theoretical predictions that there should be a large
reservoir of hot gas left over from the formation
of the Local Group (e.g., Cen \& Ostriker 1999).
However, further FUV absorption line measurements and
additional X-ray observations will be required to test
this interesting idea.  

It is unlikely that the high-velocity O\,{\sc vi} is
produced by photoionization. Probably, the gas
is collisionaly ionized at temperatures of several
$10^5$ K. The O\,{\sc vi} then may be produced
in the turbulent interface regions between very hot
($T>10^6$ K) gas in an extended Galactic Corona
and the cooler gas clouds that are moving through
this hot medium
(see Sembach et al.\,2003).
Evidence for the existence of such 
interfaces also comes from the comparison of absorption 
from neutral species like O\,{\sc i} with
absorption from highly-ionized species
like O\,{\sc vi} (Fox et al.\,2003; see also Fig.\,1). 
 
\section{Summary}

Absorption line studies towards extragalactic background
sources represent an important tool to study the halo
of the Milky Way and its immediate intergalactic 
environment. Recent
studies based on data from FUSE, STIS and other
instruments unveil a complex interplay between
a number of different processes that determine 
the distribution of cool, warm, and
hot gas around the Galaxy. These processes
include
Galactic-Fountain type flows, interaction 
of the Milky Way with the Magellanic Clouds and other satellite
galaxies, and possibly infalling Local Group gas.
The measurements demonstrate that the formation
of the Milky Way has not been completed yet.

\begin{acknowledgments}
\noindent
P.R. acknowledges support from the {\it Deutsche
Forschungsgemeinschaft.} Thank's to Don and Ron for
so many inspiring ideas about how the Galaxy works !
\end{acknowledgments}

\section{Discussion}

\noindent
{\it Kuntz:} Complex M was the first HVC complex to be detected in absorption, and
thus its distance is known to be less than $\sim4$ kpc. What limits on 
O\,{\sc vi} can be placed in this direction ?
\\

\noindent
{\it Richter:} 
The information on O\,{\sc vi} Complex M is relatively sparse. The FUSE spectrum
of TON\,1187 - the only Complex M sight line included in the FUSE O\,{\sc vi} survey -
shows no evidence for O\,{\sc vi} absorption associated with Complex M.
\\

\noindent
{\it Hurwitz:} Do there exist any FUSE observations of adjacent sight line that
can be used establish an unambiguous lower limit to the distance of 
O\,{\sc vi} absorbing gas ?
\\

\noindent
{\it Richter:}
O\,{\sc vi} exists also in the disk of the Milky Way and therefore
can be observed also toward nearby stars. The difficulty for us is
to separate O\,{\sc vi} absorption that occurs in the Milky Way
disk from absorption that is produced in the halo or disk/halo interface,
as O\,{\sc vi} absorption is broad and the velocity information is
{\it not} unambiguous.
\\

\noindent
{\it Konz:} You mentioned that there is no correlation between H\,{\sc i}
emission and O\,{\sc vi} absorption. Did you check for an anticorrelation ?
\\

\noindent
{\it Richter:}
Yes, we checked that. There is neither a correlation nor an anticorrelation
between O\,{\sc vi} and H\,{\sc i} seen in the data.
\\

\end{document}